\documentclass[conference]{IEEEtran}
\IEEEoverridecommandlockouts
\usepackage{cite}
\usepackage{amsmath,amssymb,amsfonts,mathrsfs,amsthm}
\usepackage{algorithmic}
\usepackage{graphicx}
\usepackage{textcomp}
\usepackage{xcolor}
\usepackage{subcaption}
\usepackage{makecell}
\usepackage{siunitx} 

\usepackage[left=0.625in,top=0.73in,bottom=1.0in,right=0.625in]{geometry}

\newtheorem*{remark}{Remark}
\def\BibTeX{{\rm B\kern-.05em{\sc i\kern-.025em b}\kern-.08em
    T\kern-.1667em\lower.7ex\hbox{E}\kern-.125emX}}

\begin{document}

\title{Battery Degradation-Aware Task Scheduling for Space Computing Power Networks}
\title{Battery Tax of Space Computing: Quantifying LEO Battery Aging via Physics-Driven Modeling}
\title{Unseen Cost of Space Computing: Quantifying LEO Battery Aging via Physics-Driven Modeling}

\author{
    Li Zeng$^{*}$, Jingyang Zhu$^{\dag}$, Zixin Wang$^{*}$, Yuanming Shi$^{\dag}$, Khaled B. Letaief$^{*}$
    \thanks{The work of Yuanming Shi was supported in part by the National Natural Science Foundation of China under Grant 62522117.
    This work was supported in part by the Hong Kong Research Grants Council under the Area of Excellence (AoE) Scheme Grant No. AoE/E-601/22-R.
}
    \\[0.5ex]
    \normalsize{$^{*}$Dept. of Electronic and Computer Engineering, The Hong Kong University of Science and Technology, Hong Kong}\\
    \normalsize{$^{\dag}$School of Information Science and Technology, ShanghaiTech University, Shanghai, China}\\
    \normalsize{Emails: lzengan@connect.ust.hk, \{zhujy2, shiym\}@shanghaitech.edu.cn, \{eewangzx, eekhaled\}@ust.hk}
}

\maketitle

\begin{abstract}
Low Earth Orbit (LEO) satellite constellations in the 6G era are evolving into intelligent in-orbit computational platforms, forming Space Computing Power Networks (SCPNs) to deliver global-scale computing services. However, the intensive computation within SCPN incurs a significant ``unseen cost'': the frequent charge-discharge cycles accelerate the physical degradation of satellites' life-limiting and high-cost batteries, thereby threatening the long-term operational viability of such a system. Existing approaches, often relying on indirect metrics like Depth of Discharge (DoD) and neglecting the complex, nonlinear degradation process of battery aging, fail to accurately quantify this cost. To address this, we introduce a high-fidelity, physics-driven model that quantitatively links computational workload parameters to the nonlinear battery degradation. Building on this model, we formulate a degradation-aware scheduling problem and analyze heuristic policies across different energy regimes. Simulations reveal that the optimal strategy should be adaptive: in solar-rich conditions, a myopic policy maximizing instantaneous solar utilization is superior, whereas under energy scarcity, a reactive policy leveraging real-time battery state significantly extends lifetime.
\end{abstract}


\section{Introduction}
The vision for next-generation communication systems is converging on the integration of sensing, communication, and computation, with Low Earth Orbit (LEO) satellite constellations emerging as a critical infrastructure to provide ubiquitous service. Satellites equipped with increasingly capable onboard computing resources, are transitioning from traditional communication relays to intelligent service processors \cite{chen_survey_2025}. This transformation is a direct response to the impracticality of downlinking immense volumes of orbit-generated data and the critical need for real-time, in-situ analytics \cite{mohney_terabytes_2020}. Exemplified by the TeraFLOP-scale capabilities of modern constellations like SpaceX's Starlink \cite{noauthor_spacex_nodate}, these advancements are enabling the formation of a cohesive {Space Computing Power Network (SCPN)} \cite{kuang2025space, wang_satellite_2023-1}. This distributed in-orbit computing fabric is poised to deliver global, real-time intelligent services, such as disaster monitoring and precision agriculture \cite{ruan_edge_2025}.

However, unlike terrestrial systems, where power is supplied by a stable, maintainable grid, the SCPN relies on a hybrid solar-battery power system. In this setup, frequent charge–discharge cycles, particularly during orbital eclipse periods when satellites operate solely on battery power, significantly degrade the health of irreplaceable onboard batteries. This degradation shortens the operational lifespan of the constellation and substantially increases deployment and replenishment costs \cite{yang_towards_2016}. 
For instance, V2 Mini Starlink satellites, each estimated to cost approximately \$800,000, exhibit an operational lifetime of only about five years.
A key contributor to this issue is the execution of computationally intensive tasks on satellites during eclipse phases. Consequently, it is essential to establish a quantitative model that explicitly captures the relationship between computational task scheduling and battery degradation, serving as the foundation for lifetime-aware resource management in SCPNs \cite{chen_dynamic_2022, hussein_routing_2014}.

Many existing approaches treat the relationship between task scheduling and battery degradation as implicit. While some focus on energy-efficient routing \cite{yang_towards_2016, zhang_building_2025-1, chen_dynamic_2022, hussein_routing_2014} and networking \cite{liu_green_2024}, others exploring on-board computation typically model energy consumption using simplified proxies such as Depth of Discharge (DoD) \cite{liu_-orbit_2024, li_battery-aware_2024}. However, these indicators are only indirectly correlated with the complex, nonlinear electrochemical mechanisms that drive actual battery aging. Consequently, the relationship between computational task parameters (i.e., workload intensity, processor frequency, and execution duration) and the physical degradation of onboard batteries has not been adequately characterized.

In this paper, we establish, for the first time, a physics-informed model that directly links the execution of a computational task on a satellite to the physical degradation of its onboard battery. Leveraging this model, we formulate and analyze the fundamental problem of single-task scheduling, where the goal is to minimize long-term battery aging. Through the design and evaluation of several heuristic strategies, we uncover a crucial insight:
\textbf{The optimal scheduling heuristic is highly context-dependent, varying with the interplay between task power demand and available solar energy.} This work serves as a foundational step toward addressing more complex multi-task and distributed scheduling challenges in the emerging SCPN paradigm \cite{zhang_distributed_2025}.

\section{System Model}
\subsection{Network Architecture and Orbital Dynamics}
We consider a SCPN comprised of a constellation of $N$ heterogeneous LEO satellites, denoted by $\mathcal{S} = \{1, \dots, N\}$, which collectively function as a distributed computing platform. The physical topology of this network is modeled as a Walker Delta pattern, which is one of the most widely adopted architectures in modern satellite mega-constellations. This configuration consists of $P$ orbital planes, each with $S$ satellites, for a total of $N = P \times S$ satellites. Trajectory of each satellite $s$ is modeled by calculating its position vector $\vec{r}_s(t)$ in the Earth-Centered Inertial (ECI) frame. For any satellite in a circular orbit, its position is determined by:
\begin{equation} \label{eq:eci_position_vector}
\left\{
\begin{aligned}
    x(t) &= a (\cos\Omega \cos u(t) - \sin\Omega \sin u(t) \cos i) \\
    y(t) &= a (\sin\Omega \cos u(t) + \cos\Omega \sin u(t) \cos i) \\
    z(t) &= a (\sin u(t) \sin i)
\end{aligned}
\right.
\end{equation}
where $a = R_E + h$ is the semi-major axis, $n = \sqrt{\mu/a^3}$ is the mean motion, and $u(t) = u_0 + n \cdot t$ is the argument of latitude. These equations depend on a set of unique initial conditions for each satellite: its inclination $i$, Right Ascension of the Ascending Node (RAAN) $\Omega$, and initial argument of latitude $u_0$.

To initialize the entire constellation, we instantiate these parameters for each specific satellite according to the Walker Delta configuration. Let $s_{i,j}$ denote the $j$-th satellite in the $i$-th orbital plane, where $i \in \{0, \dots, P-1\}$ and $j \in \{0, \dots, S-1\}$. While all satellites share a common altitude $h$ and inclination $i$, their RAAN and initial phasing are unique. The RAAN for the $i$-th plane, $\Omega_i$, is given by the uniform planar spacing: $\Omega_i = i \cdot \frac{2\pi}{P}$.
The initial argument of latitude for satellite $s_{i,j}$, which defines its starting position within its plane, is given by:
$
    u_{0,i,j} = \left( \frac{j}{S} + \frac{i \cdot F}{N} \right) 2\pi,
$
where $F \in \{0, \dots, P-1\}$ is the integer phasing parameter between adjacent planes. By substituting $\Omega_i$ and $u_{0,i,j}$ into the general equations of motion \ref{eq:eci_position_vector}, we can compute the precise coordinates of any satellite at any given time $t$.

\subsection{Satellite On-Board Energy Harvesting Model}
The energy dynamics of a satellite are governed by the balance between harvested solar power and onboard power consumption. Energy harvesting is the satellite's primary power source, achieved by converting solar radiation into electricity via its solar panels. The generated power, denoted as $P_s^H(t)$, is highly time-variant as it directly depends on the satellite's orbital position, which was modeled in the previous section. On the other side of the balance, energy consumption consists of two main components: a constant baseline power, $P_s^{\text{operational}}$, required for basic platform operations, and a variable power draw from payloads performing tasks such as computation, communication, or sensing. The magnitude of this payload-related consumption depends on the specific scheduling decisions made within the SCPN.

In the energy harvesting process, the instantaneous harvested solar power, $P_s^H(t)$, is a function of the solar irradiance, panel properties, and two critical geometric factors derived from the orbital model: the satellite's eclipse state and the cosine loss from panel orientation, which is expressed as \cite{wang2025emulating}: 
\begin{equation}
    P_s^H(t) = I_s(t) \cdot G_{\text{solar}} \cdot A_s \cdot \eta_s \cdot \cos(\theta_{\text{eff}}(t))
    \label{eq:harvesting_main}
\end{equation}
where $G_{\text{solar}} \approx \SI{1361}{\watt\per\meter\squared}$ is the solar constant. The two time-varying components, $I_s(t)$ and $\cos(\theta_{\text{eff}}(t))$, are detailed below.

\subsubsection{Eclipse Condition}
The satellite cannot harvest energy when it is in the Earth's geometric shadow. The binary eclipse indicator, $I_s(t) \in \{0, 1\}$, is 0 during an eclipse and 1 otherwise. An eclipse occurs when two conditions are met: 1) the satellite is on the night side of the Earth, and 2) its perpendicular distance from the Sun-Earth line is less than Earth's radius. Let $\vec{r}_{\text{sun}}$ be the vector to the Sun. The indicator is formally:
\begin{equation}
    I_s(t) = 
    \begin{cases} 
      0 & \text{if } \left( \vec{r}_s(t) \cdot \vec{r}_{\text{sun}} < 0 \right) \land \\
        & \left( \|\vec{r}_s(t)\|^2 - (\vec{r}_s(t) \cdot \hat{u}_{\text{sun}})^2 < R_E^2 \right) \\
      1 & \text{otherwise}
    \end{cases}
\end{equation}
where $\hat{u}_{\text{sun}}$ is the unit vector towards the Sun.

\begin{figure}
    \centering
    \includegraphics[width=1\linewidth]{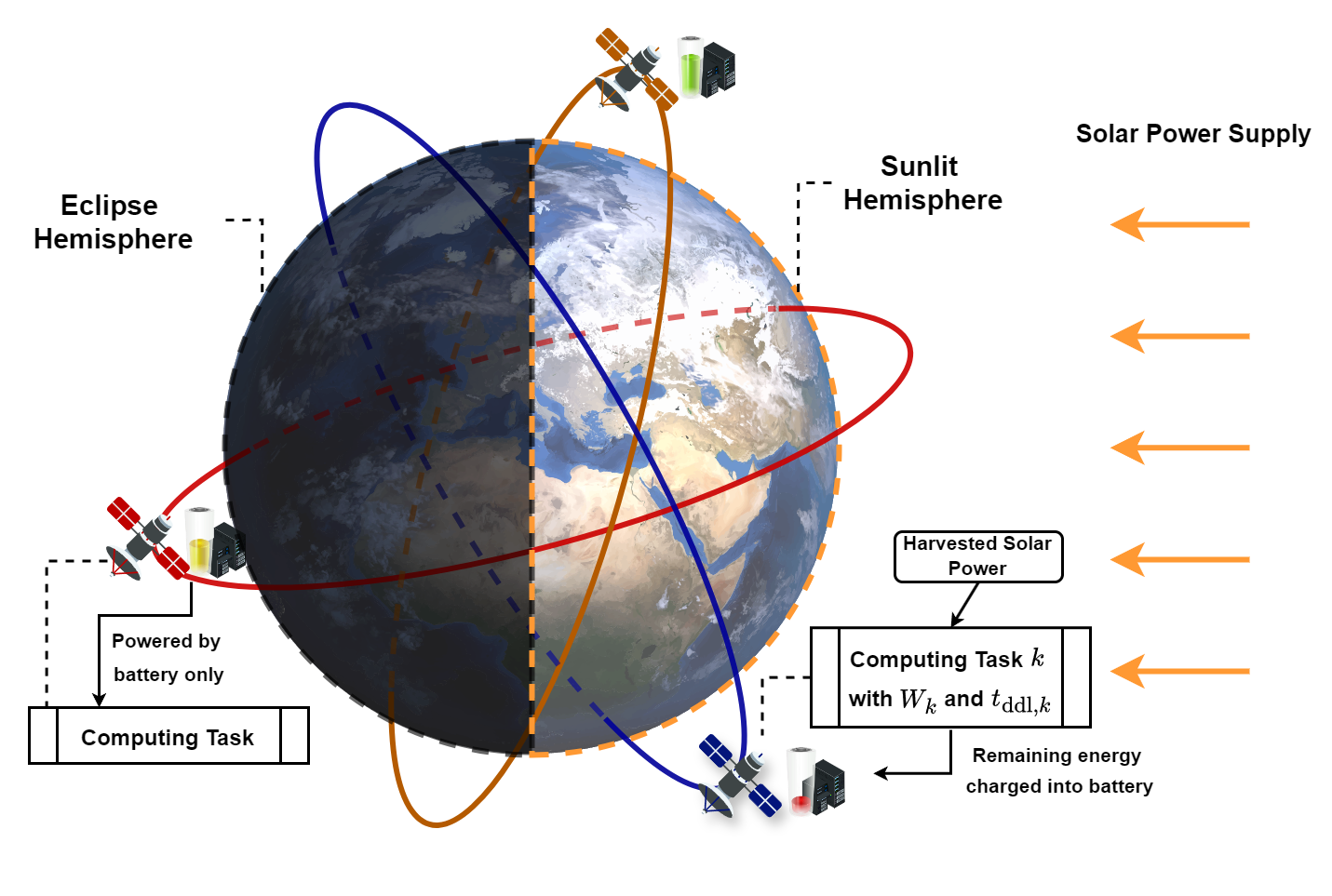}
    \caption{System model overview. A satellite's power source for a computational task depends on its orbital position. In the sunlit hemisphere, it primarily uses harvested solar power, whereas in the eclipse hemisphere, it relies solely on its battery, which accelerates degradation.}
    \label{fig:placeholder}
    \vspace{-0.5cm}
\end{figure}

\subsubsection{Panel Orientation and Cosine Loss}
We assume solar panels are fixed and point radially away from the Earth's center. The panel's normal vector is thus aligned with the satellite's position vector, $\hat{n}_p(t) = \vec{r}_s(t)/a$. The power harvested is proportional to the cosine of the angle between this normal and the incoming sunlight. This cosine factor is computed via the dot product:
\begin{equation}
    \cos(\theta_{\text{eff}}(t)) = \max\left(0, \frac{\vec{r}_s(t) \cdot \hat{u}_{\text{sun}}}{a}\right)
\end{equation}
With all components defined, Eq. \eqref{eq:harvesting_main} provides a complete, physics-based model for the harvested power at any time $t$.

Having the satellite's orbital and energy dynamics established, we are now equipped to model the long-term battery degradation of energy consumption.

\section{Battery Degradation Model for Computation Tasks}
While the previous section modeled the instantaneous power flow, this section introduces our core contribution: a physics-based model to quantify the cumulative and irreversible damage, i.e., battery degradation, caused by these power dynamics.
\subsection{Battery Cycle Life Model}
For Lithium-ion cells commonly used in satellites, there exists an established empirical relationship between the total number of cycles a battery can endure, $L$, and the constant DoD, $d$, applied in each cycle, which is expressed as \cite{yang_towards_2016}:
\begin{equation}
    \log_{10}(L) + \sigma \cdot d = \epsilon,
    \label{eq:cycle_life_base}
\end{equation}
where $\sigma$ and $\epsilon$ are battery-specific constants derived from experimental data. This equation captures the fundamental principle that deeper discharge cycles inflict more stress on the battery's chemistry, leading to an accelerated reduction in its overall lifespan.

\subsection{Instantaneous Degradation Rate}
In a realistic satellite operation scenario, the DoD is not constant but varies continuously. To handle this dynamic nature, we must derive an instantaneous degradation rate function, denoted as $f(d)$, which represents the rate of life consumed at a specific DoD level. Once this instantaneous rate is established, it can be integrated over any arbitrary DoD trajectory to quantify the precise battery degradation from any dynamic process, which is the core of our task-level degradation model.

The derivation, following the methodology in the literature \cite{yang_towards_2016}, begins by relating the cycle life $L$ at an arbitrary DoD $d$ to a baseline cycle life, $\hat{L}$, defined at $d=1$ (i.e., 100\% DoD). From Eq. \eqref{eq:cycle_life_base}, this relationship is:
\begin{equation}
    L = \hat{L} \cdot 10^{\sigma(1-d)}.
\end{equation}
The fraction of total life consumed by a single cycle of depth $d$ can be seen as the ratio of the baseline life to the life at that depth, which is $\hat{L}/L = 10^{\sigma(d-1)}$. By assuming the average degradation rate is uniform across the discharge, we can define the total integrated life consumption, $g(d)$, for a single discharge from 0 to $d$ as the depth multiplied by this fractional cost:
\begin{equation}
    g(d) = \int_0^d f(x)\mathrm{d}x = d \cdot 10^{\sigma(d-1)}.
    \label{eq:integrated_consumption}
\end{equation}
Then, we can find the instantaneous rate $f(d)$ by differentiating $g(d)$:
\begin{align}
    f(d) = \frac{\mathrm{d}}{\mathrm{d}d} \left( d \cdot 10^{\sigma(d-1)} \right) = 10^{\sigma(d-1)} \cdot (1 + \sigma \ln(10) \cdot d).
    \label{eq:instantaneous_rate}
\end{align}
This function $f(d)$ quantifies the marginal aging cost for operating the battery at any given DoD, providing the key tool for the analysis in the next subsection.

\subsection{Total Degradation from a Computational Task}
The final step is to apply this model to quantify the total degradation incurred by executing a specific computational task $k$. A task is characterized by its total workload, $W_k$ (e.g., in CPU cycles). To execute this task, satellite $s$ can dynamically adjust its processor frequency, $f_{s,k}$. This choice creates a critical trade-off: executing at a higher frequency reduces the task's duration, $\tau_{s,k} = W_k / f_{s,k}$, but dramatically increases its instantaneous power draw, as the dynamic power of a CMOS processor is approximately proportional to the cube of its frequency \cite{he_intelligent_2025}:
\begin{equation}
    P_{s,k}^{\text{task}} = c_s \cdot f_{s,k}^3,
\end{equation}
where $c_s$ is a constant specific to the processor's architecture.

It is worth noting that the operation of the satellite's power system follows a clear logic: harvested solar energy is first used to directly power the satellite's operational and payload needs \cite{liu_-orbit_2024}. If the harvested power exceeds the total demand, the surplus energy is used to charge the battery. Conversely, if the demand is greater than the harvested power, the battery must discharge to cover the deficit. This means battery degradation, which is primarily caused by discharging, only occurs when the total power consumption exceeds the instantaneous harvested power. This logic leads to a twofold scheduling preference: first, executing tasks during periods of high energy harvesting, and second, performing tasks at lower power levels when deadlines permit, as both strategies help minimize battery discharge.

The total life consumed, $L_{s,k}$, during a task is therefore the integral of the instantaneous degradation rate $f(d)$ multiplied by the rate of DoD increase. This can be expressed as an integral over the task's duration $\tau_{s,k}$, which starts at time $t$:
\begin{equation}
    L_{s,k} = \int_{t}^{t+\tau_{s,k}} f(d_s(t')) \cdot \max\left(0, \frac{\mathrm{d}}{\mathrm{d}t'}d_s(t')\right) \mathrm{d}t'.
    \label{eq:degradation_integral}
\end{equation}
The term $\max\left(0, \frac{\mathrm{d}}{\mathrm{d}t'}d_s(t')\right)$ acts as a switch, ensuring that the integral only accumulates value when the battery is actively discharging ($\frac{\mathrm{d}}{\mathrm{d}t'}d_s(t') > 0$). The rate of change of DoD is driven by the net power flow defined by the energy logic above:
\begin{equation}
    \frac{\mathrm{d}}{\mathrm{d}t'}d_s(t') = \frac{ (P_{s,k}^{\text{task}} + P_s^{\text{operational}}) - P_s^H(t') }{B_s^{\text{capacity}}}.
\end{equation}
By substituting the expressions for $f(d_s(t'))$ and $\frac{\mathrm{d}}{\mathrm{d}t'}d_s(t')$ into Eq. \eqref{eq:degradation_integral}, we obtain a complete and physically-accurate model for the battery life degradation caused by scheduling a given computational task. This integral, while complex, can be solved numerically to find the precise aging cost.
\section{Standard Formulation of Single Task Scheduling with Degradation }
We consider a scenario where at a given time $t_{\text{now}}$, a single computational task $k$, characterized by a workload $W_k$, arrives and needs to be executed. This task must be completed before a specified deadline, $t_{\text{ddl},k}$. For each satellite $s \in \mathcal{S}$, its current state (including position, velocity, and battery level $B_s(t_{\text{now}})$) and all its physical parameters are known.
We in this paper target at determining the optimal scheduling decision, which involves selecting a single satellite to execute the task, a start time for the execution, and a processor frequency, to minimize the resulting battery degradation while satisfying all operational constraints.
\footnote{Note that we hereby focus exclusively on the computation delay, abstracting away transmission latencies. This simplification is justified as computation time often overwhelms communication latency for the intensive workloads considered, and it allows us to better isolate the core trade-offs of the task allocation itself.}

The single-task scheduling problem can be formulated as a mixed-integer nonlinear programming (MINLP) problem. The goal is to choose an optimal triplet of a satellite assignment, a start time, and a processing frequency. For a given task $k$ and the set of satellites $\mathcal{S}$, we define the following decision variables: a binary assignment variable $x_s \in \{0, 1\}$, where $x_s=1$ if the task is assigned to satellite $s$; a continuous start time $t_{\text{start},s} \in \mathbb{R}^+$; and a continuous processing frequency $f_{s,k} \in \mathbb{R}^+$. The complete optimization problem, which we denote as $\mathscr{P}_1$, is formulated as follows:
\begin{subequations} \label{prob:main}
\begin{align}
    \operatorname*{minimize}_{x_s,\, t_{\text{start},s},\, f_{s,k}} \quad & \sum_{s \in \mathcal{S}} x_s \cdot L_{s,k}(t_{\text{start},s}, f_{s,k}) \label{prob:obj} \\
    \operatorname{subject\ to} \quad & \sum_{s \in \mathcal{S}} x_s = 1, \label{prob:c_assign} \\
    & x_s \in \{0,1\}, && \forall s \in \mathcal{S}, \label{prob:c_binary} \\
    & t_{\text{start},s} \geq t_{\text{now}}, &&\forall s \in \mathcal{S}, \label{prob:c_start} \\
    & t_{\text{start},s} + \frac{W_k}{f_{s,k}} \leq t_{\text{ddl},k}, &&\forall s \in \mathcal{S}, \label{prob:c_deadline} \\
    & f_{\min,s} \leq f_{s,k} \leq f_{\max,s}, &&\forall s \in \mathcal{S}, \label{prob:c_freq} \\
    & B_s(t') \geq B_{\min,s}, &&\forall s, \forall t' \in \mathcal{T}_{s,k}. \label{prob:c_battery}
\end{align}
\end{subequations}
The objective function \eqref{prob:obj} minimizes the total battery degradation, where $L_{s,k}(\cdot)$ is the degradation function from Eq. \eqref{eq:degradation_integral} and $\mathcal{T}_{s,k} = [t_{\text{start},s}, t_{\text{start},s} + W_k/f_{s,k}]$ is the task execution interval. The optimization is subject to several operational constraints. Constraint \eqref{prob:c_assign} ensures the task is assigned to exactly one satellite. Constraints \eqref{prob:c_start} and \eqref{prob:c_deadline} enforce the task's time window. Constraint \eqref{prob:c_freq} guarantees the chosen frequency is within the processor's hardware limits. Finally, constraint \eqref{prob:c_battery} is a critical safety constraint to protect the battery from deep discharge. The non-convex nature of the objective and constraints makes $\mathscr{P}_1$ intractable to solve optimally in real-time, necessitating the development of effective heuristic approaches.

\begin{remark}
Our degradation model in Eq. \eqref{eq:degradation_integral} provides two critical advantages over simpler, DoD-based metrics. First, it is \textbf{path-dependent}: it accurately captures the fact that a rapid, high-power discharge inflicts more damage than a slow, low-power discharge, even if both result in the same final DoD. This is crucial for evaluating CPU frequency scaling decisions. Second, it is \textbf{state-aware}, as the instantaneous degradation rate $f(d)$ is non-linearly dependent on the battery's current state (DoD). A scheduling decision that pushes a nearly-full battery from 10\% to 20\% DoD has a much lower aging cost than one that pushes a nearly-empty battery from 80\% to 90\% DoD. Traditional DoD metrics, which only consider the total change in charge, are blind to these critical, non-linear effects.
\end{remark}
\section{Simulation Study}
In this section, we conduct extensive simulations to evaluate the performance of several heuristics based on different scheduling strategies.
\subsection{Simulation Setup}
We establish a heterogeneous simulation environment based on the parameters detailed in Table \ref{tab:sim_params}.

\subsubsection{Constellation and Satellite Parameters}
We simulate a Walker Delta constellation with 12 orbital planes and 25 satellites per plane (300 in total), orbiting at an altitude of \SI{550}{\kilo\meter} with a \SI{53}{\degree} inclination. Key physical and power parameters for each satellite, such as its solar panel area, are randomly sampled from uniform distributions.

\begin{table}[t]
\vspace{0.2cm}
\centering
\caption{Key Simulation Parameters}
\label{tab:sim_params}
\begin{tabular}{ll}
\Xhline{1pt}
\textbf{Parameter} & \textbf{Value / Distribution} \\
\Xhline{0.5pt}
Battery Capacity ($B_s^{\text{capacity}}$) & $\SI{1200}{\watt\hour}$ \\
Solar Panel Area ($A_s$) & $U[\SI{3}{\meter\squared}, \SI{15}{\meter\squared}]$ \\
Panel Efficiency ($\eta_s$) & $U[0.05, 0.15]$ \\
Operational Power ($P_s^{\text{operational}}$) & $U[\SI{50}{\watt}, \SI{100}{\watt}]$ \\
Initial State-of-Charge & $U[20\%, 95\%]$ \\
Minimum State-of-Charge & $20\%$ \\
Processor Frequency Range & [\SI{1}{\giga\hertz}, \SI{4}{\giga\hertz}] \\
Degradation Constant ($\sigma$) & $0.8$ \\
Processor Constant ($c_s$) & $10^{-26}$ \\
Workload ($W_k$) & $U[\num{1e11}, \num{1e12}]$ cycles \\
Time Budget ($\Delta t$) & $U[\SI{25}{\second}, \SI{1000}{\second}]$ \\
\Xhline{1pt}
\end{tabular}
\end{table}

\subsubsection{Task Generation and Experiment Design}
We evaluate the heuristics across a wide range of scenarios. For our overall performance comparison, we simulate the arrival of 1000 tasks at random times over 5400s (approximately a full orbital period). For our parameter sweep experiments, we fix one task parameter while varying another, sampling 1000 different arrival times for each point to obtain statistically stable average trends.

\subsubsection{Benchmarks}
We investigate several heuristic policies as practical approaches to the problem $\mathscr{P}_1$, which focus on the most critical decision: the selection of the satellite to execute the task.
{
\begin{enumerate}
    \item[$\bullet$] \textit{Random Selection:} it selects a satellite uniformly at random from the pool of available and feasible candidate satellite.
    
    \item[$\bullet$]  \textit{DoD-First Selection:} This greedy heuristic selects the satellite with the lowest DoD, or equivalently, the highest current battery charge level, $B_s(t_{\text{now}})$. By choosing the satellite with the most stored energy, this policy aims to keep the battery operating in a less damaging, lower-degradation regime.

    \item[$\bullet$]  \textit{Min-Power-Deficit Selection:} This represents another greedy policy, but one that focuses on the current power flow rather than the stored energy. It selects the satellite with the largest instantaneous power surplus, calculated as $P_s^H(t_{\text{now}}) - P_s^{\text{operational}}$. 
    The rationale is to choose the satellite that can best cover the additional task power with currently harvested energy, thereby minimizing the immediate need to draw from the battery.

    \item[$\bullet$]  \textit{Min-Net-Energy-Cost Selection:} This advanced heuristic leverages predictive information to identify satellites expected to incur the minimum net energy drain. For a potential execution plan on each satellite (e.g., starting at $t_{\text{now}}$ at the lowest feasible frequency), it numerically integrates the predictable future harvesting curve $P_s^H(t)$ to calculate the net energy cost, defined as total energy consumed minus total energy harvested. The selection logic then prioritizes self-sustainability: any satellite with a non-positive net energy cost (indicating an energy surplus) is considered optimal and added to a candidate list. A final satellite is then chosen uniformly at random from this list. If no satellite achieves an energy surplus, the heuristic defaults to selecting the single satellite with the minimum positive net energy cost. This method uses predictable future state, which can be pre-calculated and stored in look-up tables for efficiency.

\end{enumerate}
}
Once a satellite is selected by any of the above heuristics, a common and deterministic local policy is employed to set the execution parameters ($t_{\text{start},s}$ and $f_{s,k}$). This policy follows a simple energy-saving principle: to minimize instantaneous power draw ($P^{\text{task}} \propto f^3$), it finds the lowest possible processor frequency that allows the task to complete exactly at its deadline ($t_{\text{ddl},k}$) while satisfying all hardware and battery safety constraints. If no such frequency exists, the task is deemed infeasible for the selected satellite. By applying this uniform intra-satellite strategy, we ensure a fair and consistent basis for comparing the performance of the primary selection heuristics, which will be evaluated in the next section.

\subsection{Results and Discussion}
We now present and analyze the performance of the four scheduling heuristics.

\subsubsection{Impact of Energy Availability}
\begin{figure}[t!]
    \centering
    \begin{subfigure}{\linewidth}
        \centering
        \includegraphics[width=0.9\linewidth]{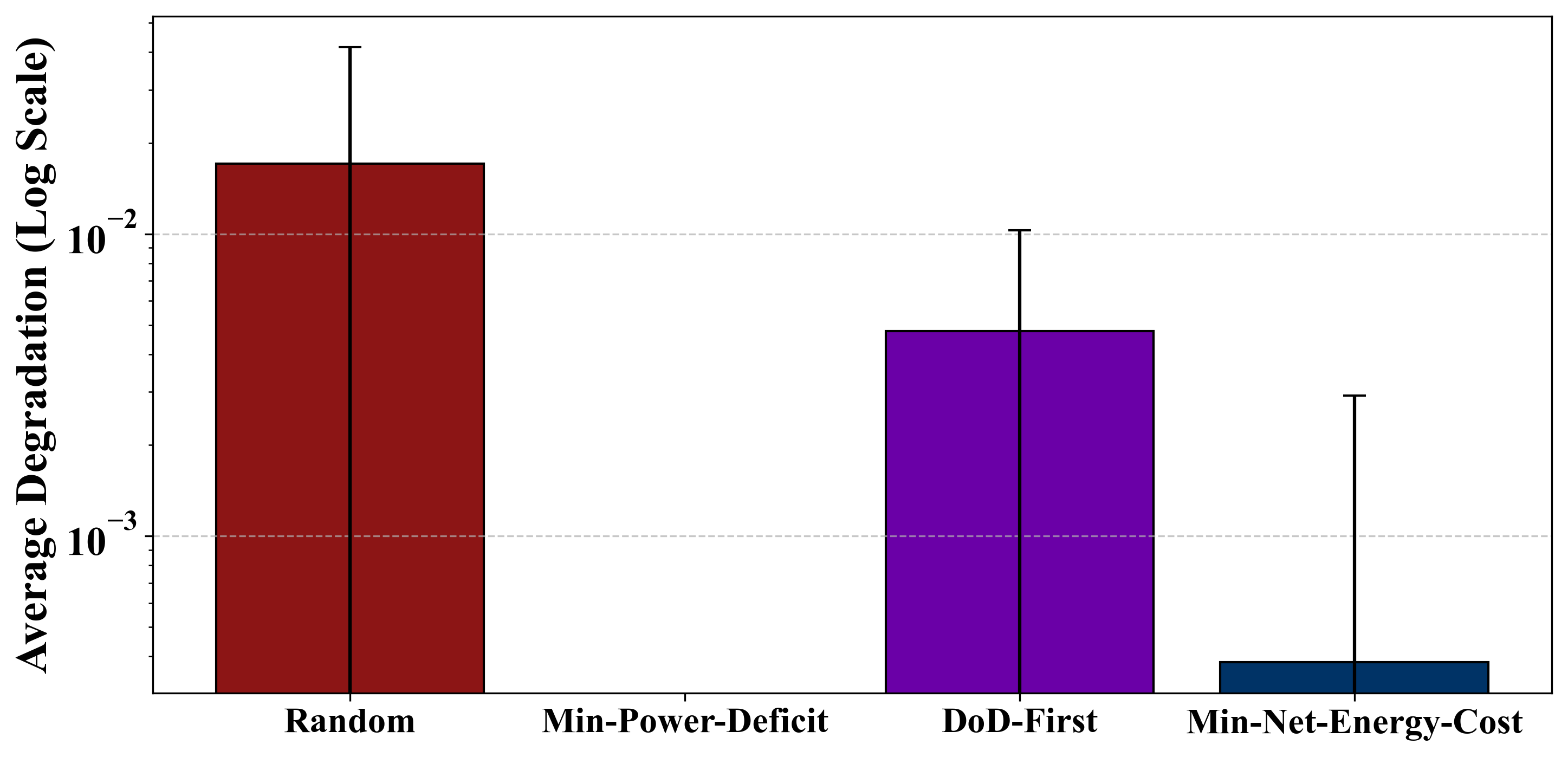}
        \caption{Performance in an energy-rich environment.}
        \label{fig:sub_high_harvest}
    \end{subfigure}
    \vfill 
    \begin{subfigure}{\linewidth}
        \centering
        \includegraphics[width=0.9\linewidth]{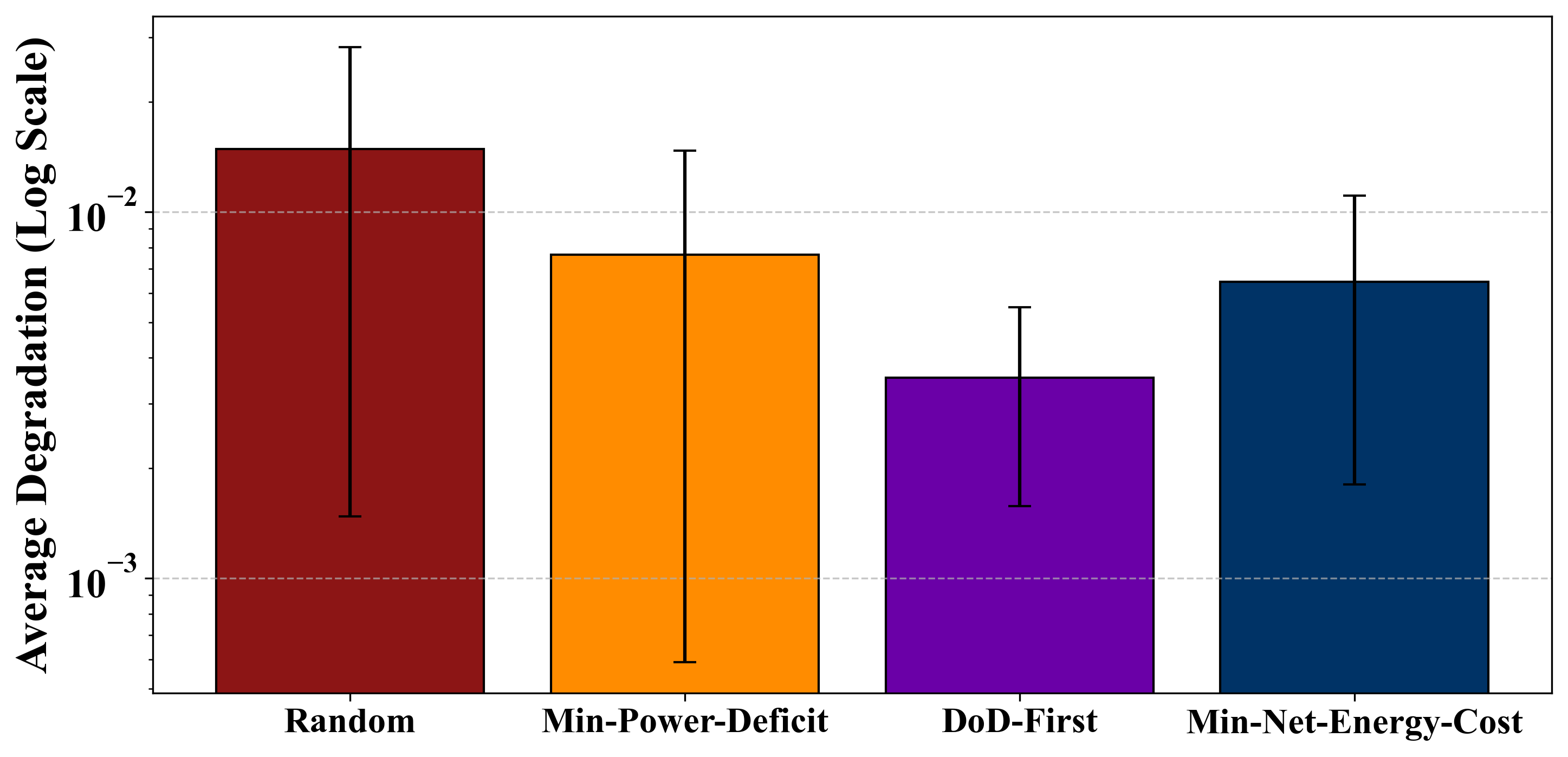}
        \caption{Performance in an energy-constrained environment.}
        \label{fig:sub_low_harvest}
    \end{subfigure}
    \caption{Average degradation cost under two different energy harvesting regimes.}
    \label{fig:overall_perf}
    \vspace{-0.6cm}
\end{figure}

The performance of the scheduling heuristics is highly dependent on the energy regime, specifically the relationship between the task's power consumption and the satellite's energy harvesting capability. We illustrate this through two contrasting scenarios, with the results summarized in Fig. \ref{fig:overall_perf}.

Fig. \ref{fig:sub_high_harvest} presents an energy-rich environment, where we set the panel efficiency range to be $U[0.1,0.3]$ and solar power is generally sufficient to cover the task's cost. Here, the \textit{Min-Power-Deficit} policy is strikingly effective, achieving near-zero degradation by consistently finding a satellite with enough instantaneous solar power to avoid battery usage entirely. The predictive \textit{Min-Net-Energy-Cost} also performs well but is slightly sub-optimal as its focus on minimizing net drain can still permit small battery discharges. Interestingly, the \textit{DoD-First} policy is a weaker performer in this regime. This reveals a potential trap: a satellite with the highest charge may be about to enter an eclipse, making it a surprisingly risky choice when solar power is otherwise abundant.

Conversely, Fig. \ref{fig:sub_low_harvest} depicts an energy-constrained environment where battery use is unavoidable. We set the panel efficiency range to be $U[0.012,0.024]$, making the task's power draw significantly higher than the available solar power. The performance hierarchy reverses dramatically. The \textit{Min-Power-Deficit} policy, which was previously optimal, now performs poorly, as its myopic focus on a non-existent power surplus becomes ineffective. In this battery-dependent regime, the strategies that explicitly manage the battery state prove superior. Notably, both \textit{DoD-First} and \textit{Min-Net-Energy-Cost} now clearly outperform the \textit{Min-Power-Deficit} policy. The \textit{DoD-First} heuristic emerges as the top performer, achieving both the lowest average degradation and the smallest variance. By ensuring tasks start at a less damaging, low-DoD state, it proves to be the most robust strategy when battery discharge is inevitable.
These results yield a crucial insight: the optimal scheduling strategy is not static. An ideal scheduler should be adaptive. \textbf{When a task's power requirement is low relative to the available harvesting, it should prioritize using instantaneous solar power}. \textbf{When the task is power-intensive and battery use is inevitable, it should shift to a battery-centric strategy}.

\subsubsection{Impact of Task Workload}
\begin{figure}[t!]
\centering
\includegraphics[width=1\linewidth]{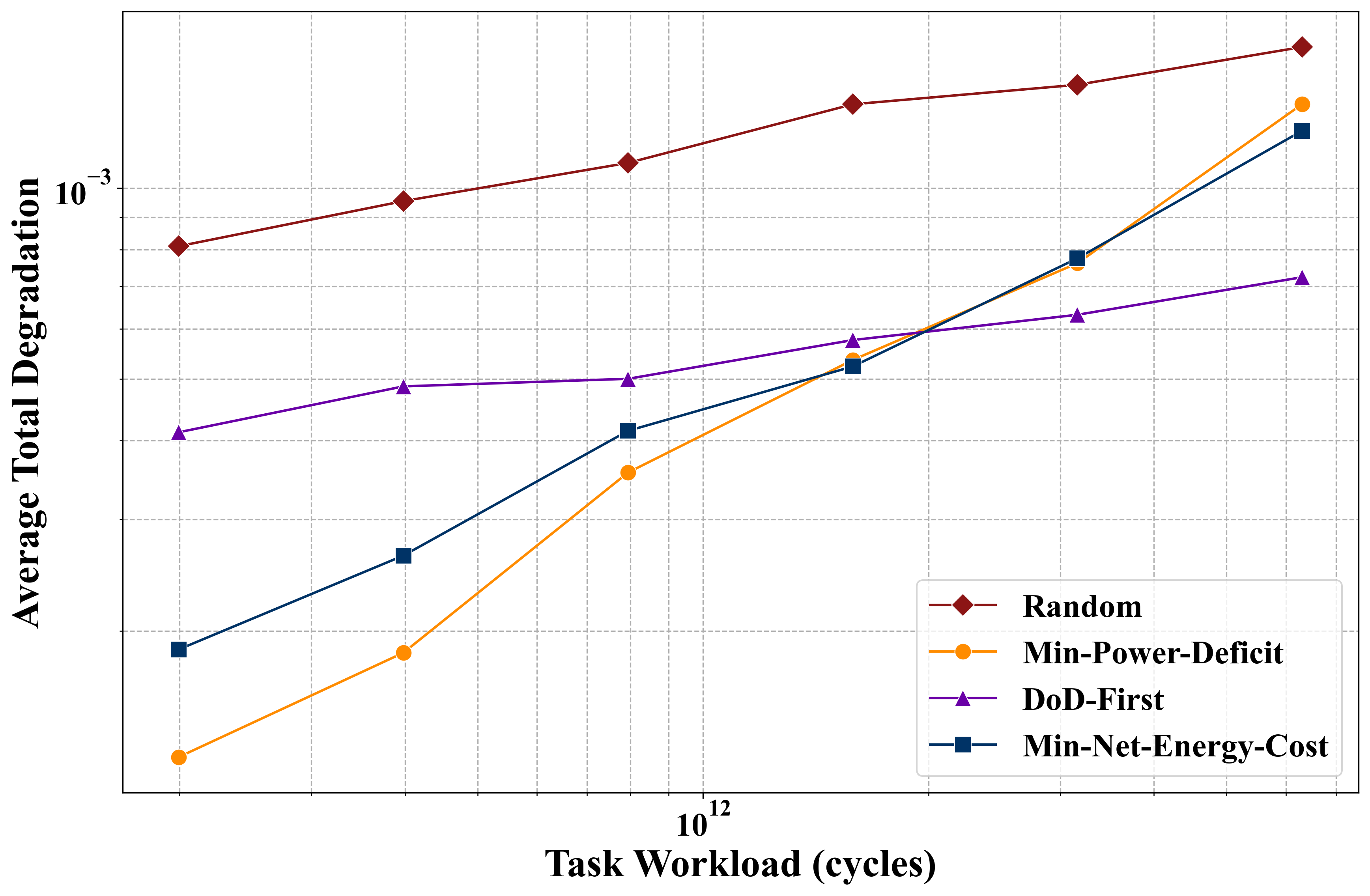}
\caption{Impact of task workload on the average degradation cost for a fixed time budget.}
\label{fig:workload_impact}
\vspace{-0.6cm}
\end{figure}

Fig. \ref{fig:workload_impact} illustrates the average degradation as the task workload increases with a fixed \SI{1500}{\second} deadline. The plot reveals a clear performance crossover. For smaller workloads (below $\approx\num{1e12}$ cycles), the myopic \textit{Min-Power-Deficit} policy excels, as the low task power can be fully sustained by satellites in direct sunlight. However, as the workload and thus the total energy demand grow, its performance degrades rapidly, and its initial power surplus becomes a poor predictor of performance over the long execution. In this high-workload regime, the battery-centric strategies become superior. The \textit{DoD-First} policy's performance remains robustly stable, proving more effective than the now-unreliable \textit{Min-Power-Deficit}. Similarly, the predictive \textit{Min-Net-Energy-Cost} demonstrates its strength by effectively managing the energy profile over the entire duration.

\subsubsection{Impact of Time Budget}
\begin{figure}[t!]
\centering
\includegraphics[width=1\linewidth]{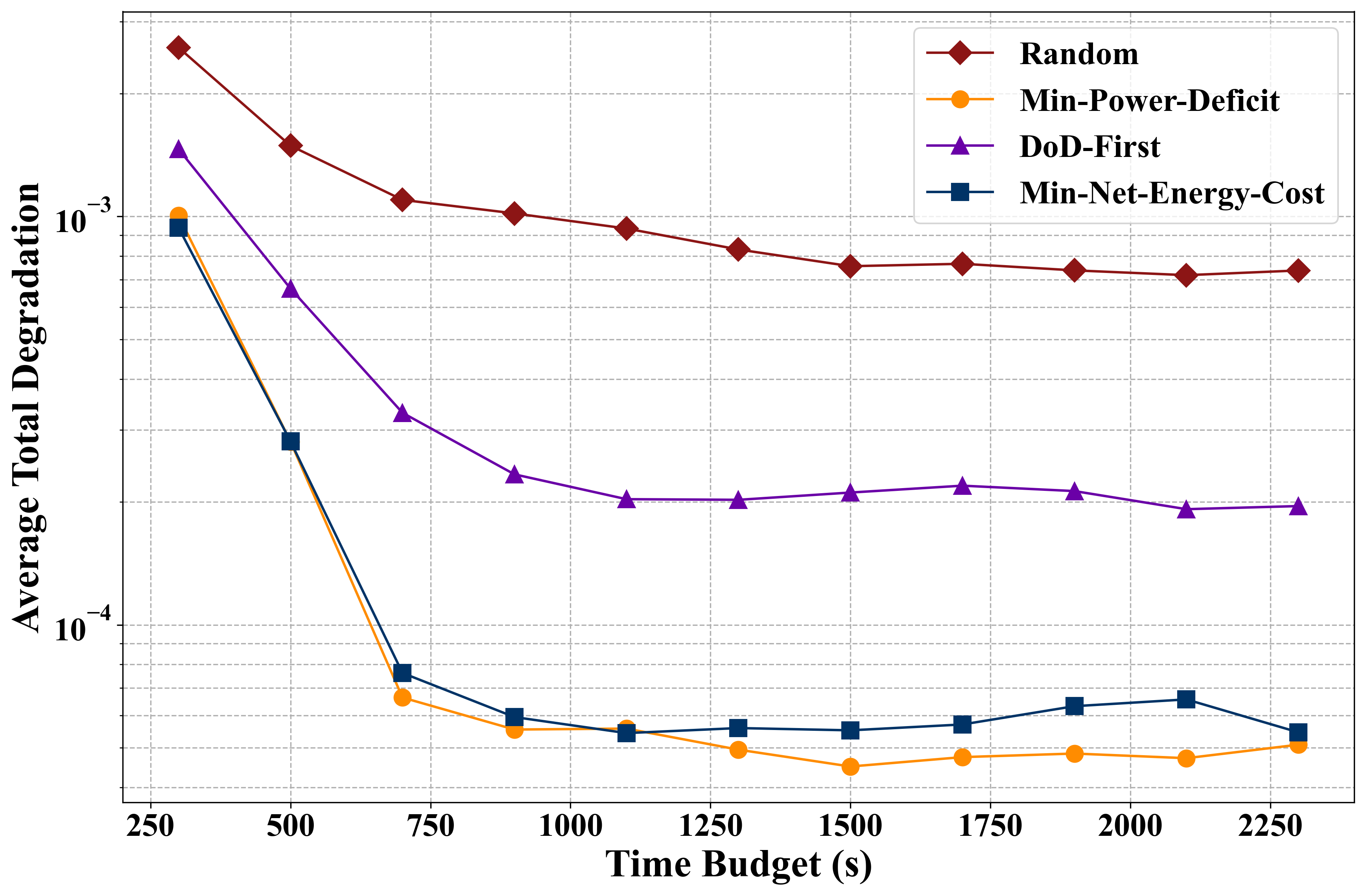}
\caption{Impact of the time budget on the average degradation cost for a fixed workload.}
\label{fig:deadline_impact}
\vspace{-0.6cm}
\end{figure}

Fig. \ref{fig:deadline_impact} illustrates the impact of the task's time budget on the average degradation for a fixed workload $10^{12}$ cycles. First, for all heuristics, increasing the time budget from very tight values (e.g., 400s to 800s) leads to a dramatic reduction in degradation. This is because the additional time allows the scheduler to select a significantly lower processor frequency, which drastically reduces the task's power draw ($P^{\text{task}} \propto f^3$) and, consequently, the stress on the battery. Second, the plot clearly again shows the superiority of the two energy-aware heuristics, \textit{Min-Power-Deficit} and \textit{Min-Net-Energy-Cost}. As the deadline becomes more flexible, they are able to find satellites where the lower-power, longer-duration task can be almost entirely powered by harvested solar energy, causing the degradation to drop to very low levels, which also demonstrates that the predictive information would be very useful.

\section{Conclusion}
In this work, we present the first physics-based quantitative model that explicitly captures the causal relationship between computational task execution and the nonlinear aging of onboard batteries in LEO satellites. Building upon this model, we formulate a battery-aging-aware single-task scheduling problem and evaluate several heuristic strategies under realistic orbital energy dynamics. Our simulations reveal a key insight: the optimal scheduling policy is inherently context-dependent: myopic, energy-efficiency-oriented strategies perform well under abundant solar power, whereas reactive policies that minimize high-current discharges are essential during eclipse phases to suppress battery degradation. This work establishes a foundational modeling framework and provides initial algorithmic insights, paving the way for future research on multi-task, distributed, and longevity-aware scheduling in sustainable SCPNs.
\bibliographystyle{IEEEtran}
\bibliography{references}

\end{document}